\documentclass[12pt]{article}
\usepackage{amssymb,amsmath,epsfig}

\begin{document}

\title{\bf On Dynamical Instability of Spherical Star in $f(R,T)$ gravity}

\author{Ifra Noureen \thanks{ifra.noureen@gmail.com} ${}^{(a)}$, M. Zubair
\thanks{mzubairkk@gmail.com; drmzubair@ciitlahore.edu.pk} ${}^{(b)}$, \\
${}^{(a)}$ Department of Mathematics,\\University of Management and Technology, Lahore, Pakistan. \\
${}^{(b)}$ Department of Mathematics,\\ COMSATS Institute of
Information Technology, Lahore, Pakistan.}

\date{}
\maketitle

\begin{abstract}
This work is based on stability analysis of spherically symmetric
collapsing star surrounding in locally anisotropic environment in
$f(R,T)$ gravity, where $R$ is Ricci scalar and $T$ corresponds to
the trace of energy momentum tensor. Field equations and dynamical
equations are presented in the context of $f(R,T)$ gravity.
Perturbation schem is employed on dynamical equations to find the
collapse equation. Furthermore, condition on adiabatic index
$\Gamma$ is constructed for Newtonian and post-Newtonian eras to
address instability problem. Some constraints on physical quantities
are imposed to maintain stable stellar configuration. The results in
this work are in accordance with $f(R)$ gravity for specific case.
\end{abstract}


\section{Introduction}

The final outcome of stellar collapse and investigations regarding
stability of compact objects is emerging as key issue in
astrophysics and gravitational theories. A star collapses when it
exhaust all its fuel and gravitational force dominates the outward
drawn pressure (Joshi and Malafarina 2011). However, the final
outcome of evolution depends on the size of compact object
undergoing collapse. The life cycle of a massive stars assuming mass
of the order $10-20$ solar masses is nominal in comparison to the
stars having relatively less mass i.e., $\approx1$ solar mass. Also,
due to high energy dissipation during collapse more massive star
tends to be more unstable because of massive radiation transport
(Hansen and Kawaler 1994; Kippenhahn and Weigert 1990).

The celestial objects are of interest only if they are stable
against fluctuations. Chandrasekhar (Chandrasekhar 1964) worked out
the dynamical instability of a spherical star, he established
instability range in the form of adiabatic index $\Gamma$ as
$\Gamma\geq\frac{4}{3}+n\frac{M}{r}$, where $M$ is the mass and $r$
stands for radius of the star. Stability analysis in General
Relativity (GR) associated with expansion free condition, isotropy,
local anisotropy, dissipation etc was presented by Herrera and his
collaborators (Chan et al. 1989, 1993, 1994). It was observed that
dissipative effects and slight change in isotropy alters subsequent
evolution considerably (Chan et al. 2000; Herrera and Santos 2003;
Herrera et al. 1989, 2004, 2012). Sharif and Abbas presented the
dynamical analysis of charged cylindrical gravitational collapse for
non-adiabatic and perfect fluid (Sharif and Abbas 2011a, 2011b).
Sharif and Azam worked out stability problem for cylindrically
symmetric thin-shell wormholes (Sharif and Azam 2013a, 2013b).

The limitations of GR on large scales urge astrophysicists towards
modified gravity explorations, they made enormous advancements to
analyze collapse and stability of gravitating objects in modified
theories of gravity. Among many modified theories of gravity, $f(R)$
gravity presents one of the elementary modification in
Einstein-Hilbert (EH) action by including higher order curvature
terms to incorporate dark source candidates. People (Cembranos et
al. 2012; Ghosh and Maharaj 2012) have discussed gravitational
collapse in modified gravity theories, Cembranos et al. investigated
collapse of self-gravitating dust particles (Cembranos et al. 2012).
The null dust non-static exact solutions have been established in
$f(R)$ gravity, constrained by constant curvature describing anti
de-Sitter background evolution (Ghosh and Maharaj 2012).

Some valuable prospects of gravitational collapse for $f(R)$ theory
of gravity are worked out in (Sharif and Kausar 2011, 2012; Kausar
2013, 2014) considering observational situations such as clustering
spectrum, cosmic microwave background and weak lensing (Carroll et
al. 2006; Bean et al. 2007; Song et al. 2007; Schmidt 2008),
concluding that inclusion of higher order curvature terms enhances
the stability range. The $f(R)$ model in the presence of
electromagnetic field assist in slowing down the collapsing
phenomenon (Kausar and Noureen 2014). Spherically symmetric collapse
of $f(R)$ gravity models is studied in (Borisov 2012) with the help
of one-dimensional numerical simulations including non-linear
coupling of scalar field. Sebastiani et al. (2013) investigate the
instabilities appearing in extremal Schwarzschild de-Sitter
background in context of modified gravity theories. Recently,
instability range of anisotropic, non-dissipative spherical collapse
is established in $f(T)$ theory (Sharif and Rani 2014). Sharif and
Abbas (2013a, 2013b) examined the dynamics of charged radiating and
shearfree dissipative collapse framed in Gauss-Bonnet gravity
theory.

Capozziello et al. (2012) analyzed the collapse and dynamics
collisionless self-gravitating systems by considering coupled
collisionless Boltzmann and Poisson equations in the context of
$f(R)$ gravity. In order to analyze the collapse and dynamics in
context of $f(R)$ gravity, the authors (Capozziello et al. 2012)
considered the coupled collisionless Boltzmann and Poisson
equations. Initially, the system is taken to be in static
equilibrium, the variation in potentials with the time transition is
measured with the help of linearly perturbed field equations.
Furthermore, Jeans wave number and Jeans mass limit is discussed on
obtention of the dispersion relation from Fourier analyzed field
equations. The Jeans instability criterion in $f(R)$ is presented by
using the dispersion relations and numerical estimation of Jeans
length in weak field limit. The numerically solved dispersion
relation is utilized for the study of interstellar medium (ISM) and
its properties.

In 2011, Harko et al. (2011) introduced $f(R,T)$ gravity theory as
another extension of GR based on matter and geometry coupling. In
$f(R,T)$, Einstein-Hilbert (EH) action is modified in a way that
matter Lagrangian includes an arbitrary function of Ricci scalar $R$
and trace of energy momentum tensor $T$. This theory can also be
conceived as generalization of $f(R)$ theory in which $T$ is
included in the action so that quantum effects or existence of some
exotic matter can be taken into account. The action in $f(R,T)$ is
written as (Harko et al. 2011)
\begin{equation}\label{1}
\int dx^4\sqrt{-g}[\frac{f(R, T)}{16\pi G}+\mathcal{L} _ {(m)}],
\end{equation}
where $\mathcal{L} _ {(m)}$ corresponds to matter Lagrangian, and
$g$ stands for the metric tensor. Different choices of $\mathcal{L}
_ {(m)}$ can be considered, each choice implies a set of field
equations for particular form of fluid.

Shabani and Farhoudi (2014) used dynamical system approach to study
the consequences of $f(R,T)$ gravity models with the help of various
cosmological parameters such as Hubble parameter, its inverse,
weight function, deceleration, snap parameters, jerk and equation of
state parameter. They explained cosmological and solar system
implications of $f(R,T)$ models and weak field limit. Sharif and
Zubair (2012a, 2012b, 2013a, 2013b) studied the laws of
thermodynamics, energy conditions and anisotropic universe models in
context of $f(R,T)$ gravity. Chakraborty (2013) formulate field
equations for homogeneous and isotropic cosmological models and
analyzed energy conditions for perfect fluid in $f(R,T)$ gravity.
Recently, dynamical instability of locally isotropic spherically
symmetric self-gravitating object is studied in (Sharif and Yousaf
2014) framed in $f(R,T)$ theory of gravity.

We aimed to find out stability range of the model under
consideration in the presence of anisotropic fluid. The perturbation
approach is used to analyze gravitational collapse in $f(R,T)$
gravity. The manuscript is arranged as: Einstein's field equations
and dynamical equations for $f(R,T)$ gravity are furnished in
section \textbf{2} that leads to the collapse equation. Section
\textbf{3} covers the discussion of stability in terms of adiabatic
index and factors affecting stability of compact objects in both
Newtonian and post-Newtonian limits. Section \textbf{4} contains
conclusion followed by an appendix.

\section{Dynamical Equations in $f(R,T)$}

The three dimensional timelike spherical boundary surface $\Sigma$ is chosen
pertaining two regions termed as interior and
exterior spacetimes. The line element for region inside $\Sigma$ is given by
\begin{equation}\label{1'}
ds^2_-=A^2(t,r)dt^{2}-B^2(t,r)dr^{2}-C^2(t,r)(d\theta^{2}+\sin^{2}\theta d\phi^{2}).
\end{equation}
The domain beyond the boundary is exterior region having line
element (Sharif and Kausar 2012)
\begin{equation}\label{25}
ds^2_+=\left(1-\frac{2M}{r}\right)d\nu^2+2drd\nu-r^2(d\theta^2+\sin^{2}\theta
d\phi^{2}),
\end{equation}
where $\nu$ corresponds to retarded time, $M$ is the total mass. It
is assumed that gravitational Lagrangian depends only on the
components of metric tensor and so corresponding energy momentum
tensor for usual matter is given by (Landau and Lifshitz 2002)
\begin{equation}\label{1'''}
 T^{(m)}_{uv}= g_{uv}\mathcal{L} _ {(m)}-\frac{2\partial\mathcal{L} _ {(m)}}{\partial g ^
 {uv}}.
\end{equation}
The variation of action (\ref{1}) with the metric $g_{uv}$
constitutes the field equations in $f(R,T)$ gravity as
\begin{eqnarray}\nonumber
&& R_{uv}f_R (R, T)-\frac{1}{2}g_{uv}f(R,
T)+(g_{uv}\square-\triangledown_u\triangledown_v)f_R (R,
T)\\\label{2} && =8\pi G T^{(m)}_{uv}-f_T (R, T)T^{(m)}_{uv}-f_T (R,
T)\Theta_{uv},
\end{eqnarray}
where $f_{R}(R,T)$ and $f_{T}(R,T)$ denote derivatives of $f(R,T)$
with respect to $R$ and $T$ respectively;
${\Box}=g^{\mu\nu}{\nabla}_{\mu}{\nabla}_{\mu}$ is the d'Alembert
operator, ${\nabla}_{\mu}$ is the covariant derivative associated
with the Levi-Civita connection of the metric tensor and
$\Theta_{{\mu}{\nu}}$ is defined by
\begin{equation}\nonumber
\Theta_{{\mu}{\nu}}=\frac{g^{\alpha{\beta}}{\delta}T_{{\alpha}{\beta}}}
{{\delta}g^{\mu{\nu}}}=-2T_{{\mu}{\nu}}+g_{\mu\nu}\mathcal{L}_m
-2g^{\alpha\beta}\frac{\partial^2\mathcal{L}_m}{{\partial}
g^{\mu\nu}{\partial}g^{\alpha\beta}}.
\end{equation}
In this study we choose $\mathcal{L}_m=\rho$ so that
$\Theta_{uv}=-2T^{(m)}_{uv}+g_{uv}\mathcal{L}_{(m)}$. Hence
Eq.(\ref{2}) takes the form
\begin{eqnarray}\nonumber
G_{uv}&=&\frac{1}{f_R}\left[(f_T+1)T^{(m)}_{uv}-\rho g_{uv}f_T
+\frac{f-Rf_R}{2}g_{uv}\right.\\\label{4}&+&\left.(\nabla_u\nabla_v-g_{uv}\Box)f_R\right].
\end{eqnarray}
Here $T^{(m)}_{uv}$ represents energy momentum tensor for usual matter describing
anisotropic fluid, given by
\begin{equation}\label{5}
T^{(m)}_{uv}=(\rho+p_{\bot})V_{u}V_{v}-p_{\bot}g_{uv}+(p_{r}-p_{\bot})\chi_{u}\chi_{v},
\end{equation}
where $\rho$ is energy density, $V_{u}$ stands for four-velocity of the fluid,
$\chi_u$ is the corresponding radial four vector, $p_r$ and $p_{\bot}$ denote
radial and tangential pressure respectively.
These physical quantities satisfy following identities
\begin{equation}\label{3}
V^{u}=A^{-1}\delta^{u}_{0},\quad
V^{u}V_{u}=1,\quad \chi^{u}=B^{-1}\delta^u_1,\quad
\chi^{u}\chi_{u}=-1.
\end{equation}
Components of field equations are
\begin{eqnarray}\nonumber
G_{00}&=&\frac{1}{f_R}\left[\rho+ \frac{f-Rf_R}{2}+
\frac{f_R''}{B^2}-\frac{\dot f_R}{A^2}\left(\frac{\dot{B}}{B}+\frac{2\dot{C}}{C}\right)\right.\\\label{f1}
&&\left.
-\frac{f_R'}{B^2}\left(\frac{B'}{B}-\frac{2C'}{C}\right)\right]
,\\\label{f2}
G_{01}&=&\frac{1}{f_R}\left[\dot{f_R}'
-\frac{A'}{A}\dot{f_R}-\frac{\dot{B}}{B}f_R'\right],\\\nonumber
G_{11}&=&\frac{1}{f_R}\left[p_r+\left(\rho+p_r\right)f_T- \frac{f-Rf_R}{2}+
\frac{\ddot{f_R}}{A^2}-\frac{\dot f_R}{A^2}\left(\frac{\dot{A}}{A}-\frac{2\dot{C}}{C}\right)
\right.\\\label{f3}
&&\left.-\frac{f_R'}{B^2}\left(\frac{A'}{A}+\frac{2C'}{C}\right)\right],
\\\nonumber
G_{22}&=&\frac{1}{f_R}\left[p_\perp+\left(\rho + p_\perp\right)f_T-
\frac{f-Rf_R}{2}+
\frac{\ddot{f_R}}{A^2}-\frac{f_R''}{B^2}-\frac{\dot
f_R}{A^2}\left(\frac{\dot{A}}{A} \right.\right.\\\label{f4}
&&\left.\left.-\frac{\dot{B}}{B}-\frac{\dot{C}}{C}\right)
-\frac{f_R'}{B^2}\left(\frac{A'}{A}-\frac{B'}{B}+\frac{C'}{C}\right)\right],
\end{eqnarray}
where dot and prime indicates the derivatives with respect to ``r''
and ``t''. Conservation laws play fundamental role in establishment
of stability range. In $f(R,T)$ gravity the energy momentum tensor
induce non-vanishing divergence. Here, we have taken into account
the conservation of full field equations i.e., Einstein tensor to
describe fluid evolution. Bianchi identities are
\begin{eqnarray}\label{bb}
&&G^{uv}_{;v}V_{u}=0,\quad
G^{uv}_{;v}\chi_{u}=0,
\end{eqnarray}
implying
\begin{eqnarray}\nonumber &&\dot{\rho}+\rho\left\{[1+f_T]\left(\frac{\dot{B}}{B}
+\frac{2\dot{C}}{C}\right)-\frac{\dot{f_R}}{f_R}\right\}+[1+f_T]
\left\{p_r\frac{\dot{B}}{B}\right.\\\label{B1}&&\left.+2p_\perp
\frac{\dot{C}}{C}\right\}+ Z_1(r,t)=0,
\\\nonumber &&
(\rho +p_r)f'_T+(1+f_T)\left\{p'_r+\rho \frac{A'}{A}+
p_r\left(\frac{A'}{A}+2\frac{C'}{C}-\frac{f'_R}{f_R}\right)
-2p_\perp\frac{C'}{C}\right\}\\\label{B2}&&+f_T\left(\rho'-\frac{f'_R}{f_R}\right)+Z_2(r,t)=0,
\end{eqnarray}
where $Z_1(r,t)$ and $Z_2(r,t)$ are given in \textbf{Appendix} as
Eqs.(\ref{B3}) and (\ref{B4}) respectively. Conservation equations
assist in describing variation from equilibrium position leading to
collapse process. The variation of physical parameters of
gravitating system with the passage of time can be observed by using
perturbation scheme as presented in the following section.

\section{$f(R,T)$ Model and Perturbation Scheme}

The gravitational field equations depicts a set of highly
complicated non-linear differential equations whose general solution
is still undetermined. We employ perturbation approach to discover
the effects of $f(R,T)$ model on the evolution of locally
anisotropic spherical star. Implementation of linear perturbation on
evolution equations leads to the pressure to density ratio i.e.,
adiabatic index describing instability range. The dynamics of
evolution can be anticipated either by following Eulerian or
Lagrangian pattern i.e., fixed or co-moving coordinates
respectively. We have chosen co-moving coordinates, all quantities
are considered to be in static equilibrium initially and then with
the time transition, the perturbed metric and material variables has
both the time and radial dependence.

In our discussion, we consider a particular model of the action
(\ref{1}) which is defined as (Sharif and Zubair 2012b)
\begin{equation}\nonumber
f(R,T)=R+\alpha R^2+\lambda T,
\end{equation}
where $\alpha$ is any real number, $\lambda$ is a coupling parameter
and $\lambda{T}$ represents correction to $f(R)$ gravity. Assuming
$0<\varepsilon\ll1$, $0<\xi\ll1$ and $0<\eta\ll1$, the perturbed
form of quantities alongwith their initial form can be arranged as
(Chan et al. 2000; Herrera and Santos 2003; Herrera et al. 1989,
2004, 2012)
\begin{eqnarray}\label{41} A(t,r)&=&A_0(r)+\varepsilon
D(t)a(r),\\\label{42} B(t,r)&=&B_0(r)+\varepsilon D(t)b(r),\\\label{43}
C(t,r)&=&rB(t,r)[1+\varepsilon D(t)\bar{c}(r)],\\\label{44}
\rho(t,r)&=&\rho_0(r)+\varepsilon {\bar{\rho}}(t,r),\\\label{45}
p_r(t,r)&=&p_{r0}(r)+\varepsilon {\bar{p_r}}(t,r),
\\\label{46}
p_\perp(t,r)&=&p_{\perp0}(r)+\varepsilon {\bar{p_\perp}}(t,r), \\\label{47}
m(t,r)&=&m_0(r)+\varepsilon {\bar{m}}(t,r), \\\label{49'}
R(t,r)&=&R_0(r)+\xi D_1(t)e_1(r),\\\label{50'}
T(t,r)&=&T_0(r)+\eta D_2(t)e_2(r),\\\nonumber
f(R, T)&=&[R_0(r)+\alpha R_0^2(r)+\lambda T_0]+\xi D_1(t)e_1(r)
[1\\\label{51'}&+&2\alpha  R_0(r)]+\eta D_2(t)e_2(r),\\\label{52'}
f_R&=&1+2\alpha R_0(r)+\xi2\alpha D_1(t)e_1(r),\\\label{52'}
f_T&=&\lambda.
\end{eqnarray}
We take Schwarzschild coordinate $C_0(r)=r$ and apply perturbation scheme on
dynamical equations i.e.,
Eqs.(\ref{B1}) and (\ref{B2}), we found that
\begin{eqnarray}\label{B1p}&&
\dot{\bar{\rho}}+\left[\frac{2e\rho_0}{Y}+\lambda_1\left\{\frac{2\bar{c}}{r}(\rho_0+2p_{\perp0})
+\frac{b}{B_0}(\rho_0+p_{r0})\right\}+YZ_{1p}\right]\dot{D}=0,
\\\nonumber
&&\lambda_1\left\{\bar{p_r}'+\bar{\rho}\frac{A'_0}{A_0}+\bar{p_r}\left(\frac{A'_0}{A_0}+\frac{2}{r}-\frac{2\alpha R'_0}{Y}\right)-\frac{2\bar{p_\perp}}{r}\right\}+\lambda\bar{\rho'}+2\alpha\ddot{D}\left[\frac{1}{A_0^2}\left(e'
\right.\right.\\\nonumber &&\left.\left.
+2e\frac{B'_0}{B_0}-\frac{b}{B_0}R'_0\right)+B_0^2(Y)\left\{\frac{e}{B_0^2Y}\right\}'\right]+D\left[\lambda_1[(\frac{a}{A_0})'(\rho_0+p_{r0})\right.\\\nonumber &&\left.-2(p_{r0}+p_{\perp0})(\frac{\bar{c}}{r})']-\frac{2\alpha}{Y}\left\{\lambda_1\left(p'_{r0}
+\rho_0\frac{A'_0}{A_0}
+p_{r0}\left(\frac{A'_0}{A_0}-\frac{2\alpha R'_0}{Y}+\frac{2}{r}\right)\right)\right\}
\right.\\\label{B2p}&&\left.+\lambda\left(e'+e[\rho'_0-\frac{2\alpha R'_0}{Y}]\right)+YZ_{2p}\right]
=0,
\end{eqnarray}
where $Y=1+2\alpha R_0$, $\lambda_1=\lambda+1$ and $Z_{1p}$,
$Z_{2p}$ are provided in appendix. It is assumed that the perturbed
quantities $D_1=D_2=D$ and $e_1=e_2=e$.

Eliminating $\bar{\dot{\rho}}$ from Eq.(\ref{B1p}) and integrating
it with respect to time provides $\bar{\rho}$ as
\begin{equation}\label{B1p'}
\bar{\rho}=-\left[\frac{2e\rho_0}{Y}+\lambda_1\left\{\frac{2\bar{c}}{r}(\rho_0
+2p_{\perp0})+\frac{b}{B_0}(\rho_0+p_{r0})\right\}+YZ_{1p}\right]D.
\end{equation}
The Harrison-Wheeler type
equation of state associating $\bar{\rho}$ and $\bar{p_r}$ depicts
describes second law of thermodynamics, written as
\begin{equation}\label{B7}
\bar{p}_r=\Gamma\frac{p_{r0}}{\rho_0+p_{r0}}\bar{\rho}.
\end{equation}
Inserting of $\bar{\rho}$ in above relation, we have
\begin{equation}\label{B8}
\bar{p}_r=-\Gamma\frac{p_{r0}}{\rho_0+p_{r0}}\left[\frac{2e\rho_0}{Y}
+\lambda_1\frac{2\bar{c}}{r}(\rho_0+2p_{\perp0})+YZ_{1p}\right]D -\Gamma p_{r0}\frac{b}{B_0}D.
\end{equation}
The expression for $\bar{p}_\perp$ can be found by applying perturbation on
field equation Eq.(\ref{f4}) and eliminating $\bar{p}_\perp$, that turns out to be
\begin{eqnarray}\label{B9}
\bar{p}_\perp=\{\frac{Y\bar{c}}{r}-2\alpha e\}\frac{\ddot{D}}{A_0^2}
-\frac{\lambda \bar{\rho}}{\lambda_1}+\left\{\left(p_{\perp0}
-\frac{\lambda}{\lambda_1}\rho_0\right)\frac{2\alpha e}{Y}+\frac{Z_3}{\lambda_1}\right\}D,
\end{eqnarray}
where $Z_3$ corresponds to effective part of the field equation and
is given in appendix as Eq.(\ref{Z3}).

Substituting $\bar{\rho}, \bar{p_r}$ and $\bar{p_\perp}$ from
Eqs.(\ref{B1p'}), (\ref{B8}) and (\ref{B9}) in second Bianchi
identity (\ref{B2p}), we obtain
\begin{eqnarray}\nonumber
&&\ddot{D}\left[\frac{2\alpha }{A_0^2Y}\left\{e'+2e\frac{B'_0}{B_0}
-\frac{b}{B_0}R'_0\right\}-2\alpha B_0^2\left\{\frac{e}{B_0^2Y}\right\}' +\frac{1}{A_0^2}\{\frac{Y \bar{c}}{r}
-2\alpha e\}\right]\\\nonumber &&+D\left[\frac{1}{Y}\left\{\lambda_1\left((\rho_0+p_{r0})\left(\frac{a}{A_0}\right)'
-2(\rho_0+p_{\perp0})\left(\frac{\bar{c}}{r}\right)'\right)
-\frac{2\alpha }{Y}\left\{\lambda\left(e'-\rho'_0
\right.\right.\right.\right.\\\nonumber &&\left.\left.\left.\left.
-\frac{2\alpha R'_0}{Y}\right)
+\lambda_1\left(e'p_{r0}
+e[p'_{r0}+\rho_0\frac{A'_0}{A_0}+p_{r0}(\frac{A'_0}{A_0}+\frac{2}{r}-\frac{2\alpha R'_0}{Y})]\right)\right\}
-\left(\lambda\right.\right.\right.\\\nonumber &&\left.\left.\left.
+\lambda_1\Gamma\frac{p_{r0}}{\rho_0+p_{r0}}\right)
\left\{
\rho_0\frac{2e}{Y}+\lambda_1\left\{\frac{2\bar{c}}{r}(\rho_0+2p_{\perp0})+\frac{b}{B_0}(\rho_0
+p_{r0})\right\}+Y Z_{1p}\right\}_{,1}\right.\right.\\\nonumber &&\left.\left.
+\left\{\frac{A'_0}{A_0}+\frac{2}{r}\frac{\lambda}{\lambda_1}+\Gamma\frac{p_{r0}}{\rho_0
+p_{r0}}(\frac{A'_0}{A_0}+\frac{2}{r}-\frac{2\alpha R'_0}{Y})
+\lambda_1\left(\Gamma\frac{p_{r0}}{\rho_0+p_{r0}}\right)'\right\}\left\{\frac{2e\rho_0}{Y}
\right.\right.\right.\\\nonumber &&\left.\left.\left.
+\lambda_1\left\{\frac{2\bar{c}}{r}(\rho_0+2p_{\perp0})
+\frac{b}{B_0}(\rho_0+p_{r0})\right\}+(Y)Z_{1p}\right\}+\frac{2}{r}\frac{1}{\lambda_1}Z_3\right\}
+Z_{2p}\right]=0.\\\label{B10}
\end{eqnarray}
An ordinary differential equation is retrieved from perturbed
form of Ricci scalar, written as
\begin{equation}\label{66}
\ddot{D}(t)-Z_4(r) D(t)=0.
\end{equation}
$Z_4$ is given in appendix as Eq.(\ref{rp}), all terms of
$Z_4$ are presumed in a way that all the terms remain positive.
The solution of Eq.(\ref{66}) takes form
\begin{equation}\label{68}
D(t)=-e^{\sqrt{Z_4}t}.
\end{equation}
Following subsections provide description of limiting cases appear
in stability problem in $f(R,T)$ gravity.

\subsection*{Newtonian Regime}

In this approximation, we take $\rho_0\gg p_{r0}$, $\rho_0\gg
p_{\perp0}$ and $A_0=1,~B_0=1$. Inserting these assumptions along
with Eq.(\ref{68}) in Eq.(\ref{66}), we arrive at following
stability condition
\begin{equation}\label{N}
\Gamma<\frac{Z_4Z_6+Z_7+\lambda\rho_0(Z_5+YZ_{1p(N)})_{,1}+X Z_5-
\frac{2}{r\lambda_1}Z_{3(N)}+YZ_{2p(N)}}{\lambda_1p_{r0}Z'_5
+\left\{p_{r0}\left(\frac{2\alpha R'_0}{Y}-\frac{2}{r}\right)\right\}Z_5},
\end{equation}
where
\begin{eqnarray}\nonumber&&
X=(\lambda\rho'_0+\frac{2\lambda}{r\lambda_1}), \quad Z_5 = \frac{2e}{Y}
+\lambda_1\left(\frac{2\bar{c}}{r}+b\right),\quad \quad Z_6 = -2\alpha^2b R'_0+Y\frac{\bar{c}}{r},
\\\nonumber && Z_7= \lambda_1\left[\rho_0 a'+2(p_{r0}+p_{\perp0})\left(\frac{\bar{c}}{r}\right)'\right]+
\frac{2\alpha}{Y}\left[\lambda\left(\frac{2\alpha
R'_0}{Y}-\rho'_0+e'\right) \right.\\\nonumber
&&\left.+\lambda_1\left\{p_{r0}+e[p'_{r0}+p_{r0}\left(\frac{2}{r}
-\frac{2\alpha R'_0}{Y}\right)]\right\}\right],
\end{eqnarray}
where $Z_{1p(N)}$ and $Z_{2p(N)}$ are terms of $Z_{1p}$ and $Z_{2p}$
corresponding to Newtonian era. The case when $\alpha\rightarrow0$
and $\lambda\rightarrow0$ corresponds to GR corrections and $\Gamma$
reads
\begin{equation}\label{N}
\Gamma<\frac{\frac{\bar{c}}{r}Z_4+\rho_0 a'+2(p_{r0}+p_{\perp0})\left(\frac{\bar{c}}{r}\right)'+e'
-\frac{2}{r^2}\left(\frac{\bar{c}''}{r}
+a''\right)}{2p_{r0}e'+p_{r0}\frac{4}{r}}
\end{equation}
These results reduce in $f(R)$ gravity for $\lambda\rightarrow0$.
The self-gravitating body remains stable in Newtonian regime until
the inequality for $\Gamma$ holds, for which following constraints
must be fulfilled.
\begin{equation}\nonumber
2\alpha^2 b R_0'<Y \frac{\bar{c}}{r}, \quad 2\alpha R'_0<Y, \quad
\frac{2\alpha R'_0}{Y}>\rho_0'-e'.
\end{equation}

\subsection*{Post Newtonian Regime}

In this approximation we take, $A_0=1-\frac{m_0}{r}$ and $B_0=1+\frac{m_0}{r}$ implying
\begin{eqnarray}\nonumber&&
\frac{A'_0}{A_0}=\frac{m_0}{r(r-m_0)}, \quad \frac{B'_0}{B_0}=\frac{-m_0}{r(r+m_0)},
\end{eqnarray}
substitution of above assumptions in Eq. (\ref{B10}) yields
\begin{equation}\label{PN}
\Gamma<\frac{Z_4Z_8+Z_9+\lambda\rho_0(Z_{10}+YZ_{1p(PN)})_{,1}+Z_{11} Z_{10}-
\frac{2}{r\lambda_1}Z_{3(PN)}+YZ_{2p(PN)}}{\lambda_1p_{r0}Z'_{10}
+\left\{p_{r0}\left(\frac{m_0}{r(r-m_0)}
+\frac{2\alpha R'_0}{Y}+\frac{2}{r}\right)\right\}Z_{10}},
\end{equation}
where
\begin{eqnarray}\nonumber
Z_8&=&\frac{2\alpha r^2}{(r-m_0)^2}\left\{e'-\frac{r}{r+m_0}\left(bR'_0+2e\frac{m_0}{r}\right)\right\}+
Y\left[\frac{ r^2}{(r-m_0)^2}\left\{2\alpha e
\right.\right.\\\nonumber&&\left.\left.-Y\frac{\bar{c}}{r}\right\}-\frac{2\alpha(r+m_0)^2}{r^2}
\left\{\frac{er^2}{Y(r+m_0)^2}\right\}'\right]
\\\nonumber
Z_9&=&\lambda_1\left\{\rho_0\left(\frac{ar}{r-m_0}\right)'-2(p_{r0}+p_{\perp0})\left(\frac{\bar{c}}{r}\right)'\right\}
-\frac{2\alpha}{Y}\left[(\lambda_1 p_{r0}+\lambda)e'\right.\\\nonumber&&\left.
+e\left\{\lambda_1( p'_{r0}+\frac{\rho_0m_0}{r(r-m_0)})+p_{r0}\left(\frac{2}{r}-\frac{2\alpha R'_0}{Y}\right)\right\}
-\lambda\left(\rho'_0-\frac{2\alpha R'_0}{Y}\right)\right]\\\nonumber
Z_{10}&=&\frac{2e}{Y}+\lambda_1\left(2\frac{\bar{c}}{r}+\frac{br}{r+m_0}\right), \quad
Z_{11}=\left(\frac{m_0}{r(r-m_0)}
+\frac{2\alpha R'_0}{Y}+\frac{2\lambda}{\lambda_1r}+\lambda\rho'_0\right)
\end{eqnarray}
$Z_{1p(PN)}$ and $Z_{2p(PN)}$ are terms of $Z_{1p}$ and $Z_{2p}$
corresponding to post-Newtonian era. Again inequality holds for
positive definite terms restricting physical quantities as under
\begin{eqnarray}\nonumber
\frac{r}{r+m_0}(bR'_0+\frac{2em_0}{r})<e',\quad 2\alpha
e-Y\frac{\bar{c}}{r}>\frac{(r^2-m_0^2)^2}{r^4}\left\{\frac{er^2}{Y(r+m_0)^2}\right\}'
\end{eqnarray}
The restrictions imposed on these quantities have significant impact
on stellar structure and must be fulfilled to achieve stability of
considered model.

\section{Summary and Discussion}

Modified theory can be considered as an effective candidate to
explain the issue of cosmic acceleration. In this setting, $f(R,T)$
gravity provides an alternative way without introducing an exotic
energy component. The matter geometry coupling results in existence
of extra force due to nongeodesic motion of test particles. This
work is devoted to examine the impact of $f(R,T)$ model on dynamical
instability of locally anisotropic spherical star in $f(R,T)$
gravity. The model $f(R,T)=R+\alpha R^2+\lambda T$ provides a viable
substitute to the exotic matter in our universe (Sharif and Zubair
2012b). The local anisotropy in matter configuration contribute high
dissipation of energy via heat flow, radiation transport, shearing
stresses etc. that affect stability range largely.

The highly complicated non-linear differential equations
corresponding to set of field equations can not be solved generally,
that is why we have employed perturbation approach to incorporate
this problem. Perturbed dynamical equations constitute collapse
equation that describe non-static spherical star. Perturbed second
Bianchi identity reveals that instability range of gravitating
objects has dependence on adiabatic index $\Gamma$. Moreover, the
index $\Gamma$ constitute impressions of various factors such as
tangential and radial pressure, shear, density inhomogeneity,
radiation, free streaming etc.

It is worth noticing that in comparison to $f(R)$ gravity the
dynamical equations of $f(R,T)$ gravity possess extra term of $T$
that depict the more generalized modification of GR. Further, adding
the terms of trace $T$ in EH action includes the description of
quantum effects or so-called exotic matter. The adiabatic index
$\Gamma$ admits the positive definite terms to maintain stable
stellar configuration for both Newtonian and post-Newtonian eras.
Inclusion of trace of energy momentum tenor in action (\ref{1})
results in positive addition to $\Gamma$ and so enhances the
stability of stars by slowing down the subsequent collapse.

It is concluded that our results are in agreement with the $\Gamma$
configuration found in (Sharif and Kausar 2012) for
$\lambda\rightarrow0$ presenting $f(R)$ gravity. Local isotropy of
the model can be retrieved by setting $p_r=p_\perp=p$ and the
results are well consistent with the findings of (Sharif and Yousaf
2014). The terms corresponding to GR corrections can be found by
assuming vanishing values of both $\alpha$ and $\lambda$.

The extension of present work to discuss stability, expansion free
and shear free conditions in $f(R,T)$ is submitted (Ifra and Zubair
2014a, 2014b). The work related shear free condition  Furthermore,
it is worth stressing that the pattern devised in (Capozziello et
al. 2012) can be generalized for various modified gravity theories
(Capozziello et al. 2008; Bogdanos et al. 2010; De Laurentis and
Capozziello 2011; Roshan and Abbassi 2014). To address instability
problem in an alternate and adequate way, the work on Jeans
instability criterion can be extended for $f(R,T)$ theory.  We are
planning to work out this problem (Jeans analysis in $f(R,T)$) in
our forthcoming article incorporating the comparison with
interstellar medium.

\section{Appendix}

\begin{eqnarray}\setcounter{equation}{1}\nonumber
Z_1(r,t)&=&f_R A^2\left[\left\{\frac{1}{f_R A^2}\left(\frac{f-Rf_R}{2}
-\frac{\dot{f_R}}{A^2}\left(\frac{\dot{B}}{B}+\frac{2\dot{C}}{C}\right)-\frac{f'_R}{B^2}\left(\frac{B'}{B}
-\frac{2C'}{C}\right)\right.\right.\right.\\\nonumber &&\left.\left.\left.
+\frac{f''_R}{B^2}\right)\right\}_{,0}+\left\{\frac{1}{f_RA^2B^2}\left(\dot{f'_R}-\frac{A'}{A}\dot{f_R}
-\frac{\dot{B}}{B}f_R'\right)\right\}_{,1}\right]
-\frac{\dot{f_R}}{A^2}\left\{\left(\frac{\dot{B}}{B}\right)^2\right.\\\nonumber
&&\left.+2\left(\frac{\dot{C}}{C}\right)^2
+\frac{3\dot{A}}{A}\left(\frac{\dot{B}}{B}+\frac{2\dot{C}}{C}\right)
\right\}+\frac{\ddot{f_R}}{A^2}\left(\frac{\dot{B}}{B}+\frac{2\dot{C}}{C}\right)+\frac{\dot{A}}{A}(f-Rf_R)\\\nonumber
&&-
\frac{2f'_R}{B^2}\left\{\frac{\dot{A}}{A}\left(\frac{B'}{B}-\frac{C'}{C}\right)+\frac{\dot{B}}{B}\left(\frac{2A'}{A}
+\frac{B'}{B}+\frac{C'}{C}\right)+
\frac{\dot{C}}{C}\left(\frac{A'}{A}-\frac{3C'}{C}\right)\right\}
\\\label{B3}
&&
+\frac{f_R''}{B^2}\left(\frac{2\dot{A}}{A}+\frac{\dot{B}}{B}\right)+\frac{1}{B^2}\left(\dot{f'_R}-\frac{A'}{A}\dot{f_R}
\right)\left(\frac{3A'}{A}
+\frac{B'}{B}+\frac{2C'}{C}\right),
\\\nonumber
Z_2(r,t)&=&f_RB^2\left[\left\{\frac{1}{f_RA^2B^2}\left(\dot{f_R}'-\frac{A'}{A}\dot{f_R}
-\frac{\dot{B}}{B}f_R'\right)\right\}_{,0}+\left\{\frac{1}{f_RB^2}\left(\frac{Rf_R-f}{2}
\right.\right.\right.
\\\nonumber &&\left.\left.\left.-\frac{\dot{f_R}}{A^2}\left(\frac{\dot{A}}{A}-\frac{2\dot{C}}{C}\right)-\frac{f'_R}{B^2}\left(\frac{A'}{A}
+\frac{2C'}{C}\right)+\frac{\ddot{f_R}}{A^2}\right)\right\}_{,1}
\right]+(Rf_R-f)\frac{B'}{B}
\\\nonumber
&&-\frac{\dot{f_R}}{A^2}\left\{\frac{A'}{A}\left(\frac{\dot{A}}{A}+\frac{\dot{B}}{B}\right)
+\frac{B'}{B}\left(\frac{\dot{A}}{A}-\frac{2\dot{C}}{C}\right)+\frac{2C'}{C}\left(\frac{\dot{B}}{B}-\frac{\dot{C}}{C}\right)\right\}
+\left(\frac{\dot{A}}{A}\right.\\\nonumber
&&\left.+\frac{3\dot{B}}{B}
+\frac{2\dot{C}}{C}\right)\left(\dot{f_R}'-\frac{A'}{A}\dot{f_R}
-\frac{\dot{B}}{B}f_R'\right)\frac{1}{A^2}-\frac{f'_R}{B^2}\left\{\frac{A'}{A}\left(\frac{A'}{A}+\frac{3B'}{B}\right)+
\right.\\\label{B4} &&\left.
\frac{2C'}{C}\left(\frac{3B'}{B}
+\frac{C'}{C}\right)\right\}+\frac{\ddot{f_R}}{A^2}\left(\frac{A'}{A}+\frac{2B'}{B}\right)
+\frac{f''_R}{B^2}\left(\frac{A'}{A}
+\frac{2C'}{C}\right).
\end{eqnarray}
\begin{eqnarray}\nonumber
Z_{1p}&=& 2\alpha A_0^2\left[\frac{1}{A_0^2B_0^2Y}\left\{e'-e\frac{A'_0}{A_0}-\frac{b}{B_0}R'_0\right\}\right]_{,1}
+\frac{1}{Y}\left[e-[\lambda T_0
\right.\\\nonumber && \left.-\alpha R_0^2]\left(\frac{a}{A_0}+\frac{e}{Y}\right)
+\frac{2\alpha}{B_0^2}\left\{\left(\frac{B'_0}{B_0}-\frac{2}{r}\right)\left(e'-2R'_0\left(\frac{a}{A_0}
+\frac{b}{B_0}\right)\right.\right.\right.\\\nonumber & &\left.\left.\left.
+\frac{2 \alpha e}{Y}R'_0\right)
+ R''_0 \left(\frac{2a}{A_0}+\frac{b}{B_0}\right)-2R'_0\left(\frac{b}{B_0}\left(\frac{2A'_0}{A_0}+
\frac{B'_0}{B_0}+\frac{1}{r}\right)\right.\right.\right.\\\label{Z1p} & &\left.\left.\left.+\frac{\bar{c}}{r}\left(\frac{A'_0}{A_0}-\frac{3}{r}\right)\right)
+\left(e'-e\frac{A'_0}{A_0}\right)\left(\frac{3A'_0}{A_0}+
\frac{B'_0}{B_0}+\frac{2}{r}\right)\right\}\right]
\\\nonumber
Z_{2p}&=&B_0^2Y\left[\frac{1}{B_0^2Y}\left\{e-\frac{2\alpha}{B_0^2}\left\{
\left(\frac{A'_0}{A_0}+\frac{2}{r}\right)\left(e-[\frac{2 \alpha e}{Y}+\frac{4b}{B_0}]R'_0\right)
+R'_0[\left(\frac{a}{A_0}\right)'
\right.\right.\right.\\\nonumber & &\left.\left.\left.
+\left(\frac{\bar{c}}{r}\right)']\right\}-[\lambda T_0
-\alpha R_0^2]\left(\frac{b}{B_0}+\frac{e}{Y}\right)\right\}\right]_{,1}
+bB_0Y\left[\frac{1}{B_0^2Y}\left\{\lambda T_0
-\alpha R_0^2
\right.\right.\\\nonumber &&\left.\left.
-\frac{4\alpha}{B_0^2}\left(\frac{A'_0}{A_0}+\frac{2}{r}\right)R'_0\right\}\right]_{,1}
+\frac{2\alpha}{B_0^2}\left[R''_0\left\{\left(\frac{a}{A_0}\right)'-2\left(\frac{A'_0}{A_0}
+\frac{2}{r}\right)\left(\frac{b}{B_0}+\frac{e}{Y}\right)
\right.\right.\\\nonumber &&\left.\left.+\left(\frac{\bar{c}}{r}\right)'
\right\}-R'_0\left\{\frac{A'_0}{A_0}\left[2\left(\frac{a}{A_0}\right)'+3\left(\frac{b}{B_0}\right)'\right]
+3\frac{B'_0}{B_0}\left[\left(\frac{a}{A_0}\right)'+2\left(\frac{\bar{c}}{r}\right)'\right]
\right.\right.\\\nonumber & &\left.\left.+
\frac{2}{r}\left[\left(3\frac{b}{B_0}\right)'+2\left(\frac{\bar{c}}{r}\right)'\right]\right\}+\left(\frac{2b}{B_0}R'_0
-e\right)
\left\{3\frac{B'_0}{B_0}\left(\frac{A'_0}{A_0}+\frac{2}{r}\right)
+\left(\frac{A'_0}{A_0}\right)^2
\right.\right.\\\label{Z2p} & &\left.\left.+\frac{2}{r^2}\right\}\right]+e\frac{B'_0}{B_0}-[\lambda T_0
-\alpha R_0^2]\left(\frac{b}{B_0}+\frac{2e}{Y}\frac{B'_0}{B_0}\right)
\end{eqnarray}
\begin{eqnarray}\nonumber
Z_{3}&=&\frac{Y}{B_0^2}\left[\frac{a''}{A_0}+\frac{\bar{c}''}{r}-\frac{A''_0}{A_0}
\left(\frac{a}{A_0}+\frac{2b}{B_0}\right)+\frac{A'_0}{A_0}\left\{\frac{2b}{B_0}\left(\frac{B'_0}{B_0}
-\frac{1}{r}\right)+\left(\frac{\bar{c}}{r}\right)'\right.\right.\\\nonumber & &\left.\left.
-\left(\frac{b}{B_0}\right)'\right\}
+\frac{B'_0}{B_0}\left\{\frac{2bB_0'}{rB_0}-\left(\frac{a}{A_0}\right)'-\left(\frac{\bar{c}}{r}\right)'\right\}+
\frac{1}{r}\left\{\left(\frac{a}{A_0}\right)'-\left(\frac{b}{B_0}\right)'\right\}
\right]\\\nonumber & &
-\frac{2\alpha e}{Y}\left\{\frac{\lambda T_0
-\alpha R_0^2}{2}-\frac{2\alpha}{B_0^2}\left(R'_0\left(\frac{A'_0}{A_0}-\frac{B'_0}{B_0}
+\frac{1}{r}\right)-R''_0\right)\right\}-\frac{2\alpha}{B_0^2}\left\{e''\right.\\\label{Z3} & &\left.
+\frac{2b}{B_0}R''_0+\left(\frac{A'_0}{A_0}-\frac{B'_0}{B_0}
+\frac{1}{r}\right)\left(\frac{2b}{B_0}R'_0-e'\right)\right\}
\end{eqnarray}
\begin{eqnarray}
\nonumber Z_{4}&=&-\frac{rA_0^2B_0}{br+2B_0\bar{c}}\left[
\frac{e}{2}-\frac{2\bar{c}}{r^3}
-\frac{1}{A_0B_0^2}\left\{A_0''[\frac{a}{A_0}+\frac{2b}{B_0}]-\frac{1}{B_0}\left(a'B_0'
\right.\right.\right.\\\nonumber && \left.\left.\left.
+a''+A_0'b'-A_0'B_0'[\frac{a}{A_0}+\frac{3b}{B_0}]\right)
+\frac{2}{r}\left\{a'+\bar{c}'A_0'-A_0'[\frac{a}{A_0}
\right.\right.\right.\\\nonumber &&
\left.\left.\left.+\frac{2b}{B_0}+\frac{\bar{c}}{r}]\right\}
+\frac{A_0}{r}\left\{\bar{c}''-\frac{b'}{B_0}-\frac{B_0'\bar{c}'}{B_0}+
\frac{3b}{B_0}+\frac{\bar{c}}{r}\right\}
+\frac{2}{r^2}[\bar{c}'\right.\right.\\\label{rp} &&
\left.\left.-\frac{b}{B_0}\frac{\bar{c}}{r}]\right\}\right] =0.
\end{eqnarray}

\vspace{0.25cm}

{\bf Acknowledgment}

\vspace{0.25cm}

We would like to thank the anonymous referee for constructive
comments.

\end{document}